\documentclass{article}

     \PassOptionsToPackage{numbers}{natbib}

\usepackage[final]{neurips_2021_ml4ps} 

\usepackage[utf8]{inputenc} 
\usepackage[T1]{fontenc}    
\usepackage{hyperref}       
\usepackage{url}            
\usepackage{booktabs}       
\usepackage{amsfonts}       
\usepackage{nicefrac}       
\usepackage{microtype}      
\usepackage{xcolor}         
\usepackage{graphicx}
\usepackage{float}
\usepackage{subfig}

\title{Online Bayesian Optimization for Beam Alignment in the SECAR Recoil Mass Separator}  

\author{%
  Sara A. Miskovich\thanks{Corresponding author at: \texttt{smiskov@slac.stanford.edu}, SLAC National Accelerator Laboratory, Menlo Park, 94025 California, USA. orcid=0000-0002-3302-838X} \\
  Department of Physics and Astronomy\\
  Michigan State University\\
  East Lansing, MI 48824 USA\\
  \And
  Fernando Montes\\
 Department of Physics and Astronomy\\
  Michigan State University\\
  East Lansing, MI 48824 USA\\
  \And
  Georg P. Berg\\
  Department of Physics\\
  University of Notre Dame\\
  Notre Dame, IN 46556 USA\\
      \And
  Jeff Blackmon\\
Department of Physics and Astronomy\\
Louisiana State University\\
Baton Rouge, LA 70803 USA\\
      \And
  Kelly A. Chipps\\
 Physics Division\\ 
 Oak Ridge National Laboratory\\
 Oak Ridge, TN 37831 USA\\
      \And
  Manoel Couder\\
  Department of Physics\\
  University of Notre Dame\\
  Notre Dame, IN 46556 USA\\
      \And
  Kirby Hermansen \\
  Department of Physics and Astronomy\\
  Michigan State University\\
  East Lansing, MI 48824  USA \\
      \And
  Ashley A. Hood \\
 Cyclotron Institute \\
 Texas A\&M University \\
 College Station, TX 77843 USA\\
      \And
  Rahul Jain \\
  Department of Physics and Astronomy\\
  Michigan State University\\
  East Lansing, MI 48824 USA\\
      \And
  Hendrik Schatz \\
  Department of Physics and Astronomy\\
  Michigan State University\\
  East Lansing, MI 48824 USA\\
      \And
  Michael S. Smith \\
 Physics Division \\ 
 Oak Ridge National Laboratory \\
 Oak Ridge, TN 37831 USA\\
      \And
  Pelagia Tsintari \\
 Department of Physics \\
 Central Michigan University \\
 Mt Pleasant, MI 48859 USA \\
      \And
  Louis Wagner \\
  Department of Physics and Astronomy\\
  Michigan State University\\
  East Lansing, MI 48824 USA\\
}

\begin{document}

\maketitle

\begin{abstract}
The SEparator for CApture Reactions (SECAR) is a next-generation recoil separator system at the Facility for Rare Isotope Beams (FRIB) designed for the direct measurement of capture reactions on unstable nuclei in inverse kinematics. To maximize the performance of the device, careful beam alignment to the central ion optical axis needs to be achieved. This can be difficult to attain through manual tuning by human operators without potentially leaving the system in a sub-optimal and irreproducible state. In this work, we present the first development of online Bayesian optimization with a Gaussian process model to tune an ion beam through a nuclear astrophysics recoil separator. We show that the method achieves small incoming angular deviations (0-1 mrad) in an efficient and reproducible manner that is at least 3$\times$ faster than standard hand-tuning. This method is now routinely used for all separator tuning.
\end{abstract}

\section{Introduction}
The SEparator for CApture Reactions (SECAR) is a next-generation recoil separator system at the Facility for Rare Isotope Beams (FRIB) optimized for direct studies of capture reaction rates of proton and alpha particles on short-lived proton-rich nuclei. Direct measurements provide a higher precision reaction rate estimate than otherwise possible, and are crucial to addressing open questions regarding explosive stellar scenarios including nova explosions, X-ray bursts, and supernovae. For a detailed description of the SECAR system, see \cite{Berg2018}. To achieve precision measurements, SECAR creates mass separation between nuclear reaction products (recoils) and the unreacted beam particles leaving a target. The design separation goals require stringent conditions on the incoming beam angle (ideally 0 and up to 3 mrad) at the SECAR target in order to satisfy the ultimate physics requirements SECAR aims to achieve. 

Traditionally, beam intensity measurements are the main diagnostic used to manually optimize the beam transmission and incoming angle. Optimization can be accomplished by minimizing the intensity on apertures installed in the target chamber that joins the beamline upstream of SECAR and the SECAR beamline, while maximizing the transmission through the SECAR target chamber. Trained operators manually adjust steering knobs to align the beam while monitoring intensity readings at multiple points along the beamline. This process can take up a significant amount of beam operation time. Tasks such as visual checks of tune quality can be operator dependent, and can leave the device in an irreproducible state below optimal performance. Additionally, given the relatively large angular acceptance of the aperture system in the target chamber (3.4 mrad), precise adjustments to the required angles are challenging for operators to achieve manually. A more robust solution would require an automated tune optimizer that enhances reproducibility, ensures objectivity when assessing tune quality, and operates with an efficiency that surpasses the speed of manual tuning when searching for the optimal parameters to achieve SECAR performance targets for each experiment. 

Machine learning model-dependent optimization methods have been successfully applied in other facilities to automate the tuning and controls of complex accelerators (e.g. \cite{mcintire2016,duris2020,shalloo2020}). In contrast to these prior studies, which focused on conventional and laser wakefield based acceleration of electron beams, ion beams (in particular, proton-rich isotope beams) need to be controlled in SECAR. Since SECAR is a novel complex device with a lack of previously recorded data (e.g. to train a neural network), online learning, where the model is trained incrementally as it collects individual data instances sequentially from the live separator machine, is required. Online Bayesian optimization presents a suitable choice to substitute the time consuming task of operator-dependent manual system adjustments to find a suitable tune. In this paper, we demonstrate the first application of Bayesian optimization with a Gaussian process model to the online image-based incoming ion beam alignment in the SECAR recoil separator. We show this method to be a robust and objective way to achieve small beam angular deviations (0-1 mrad) while improving on the efficiency of traditional tuning methods.

\section{Method}\label{method}
\paragraph{Experimental system}
\begin{figure}
\centering
\subfloat[][Layout of the incoming beamline (right) and the first section of SECAR (left) with input and output variables of the optimization labeled.]{\includegraphics[height=0.26\textheight]{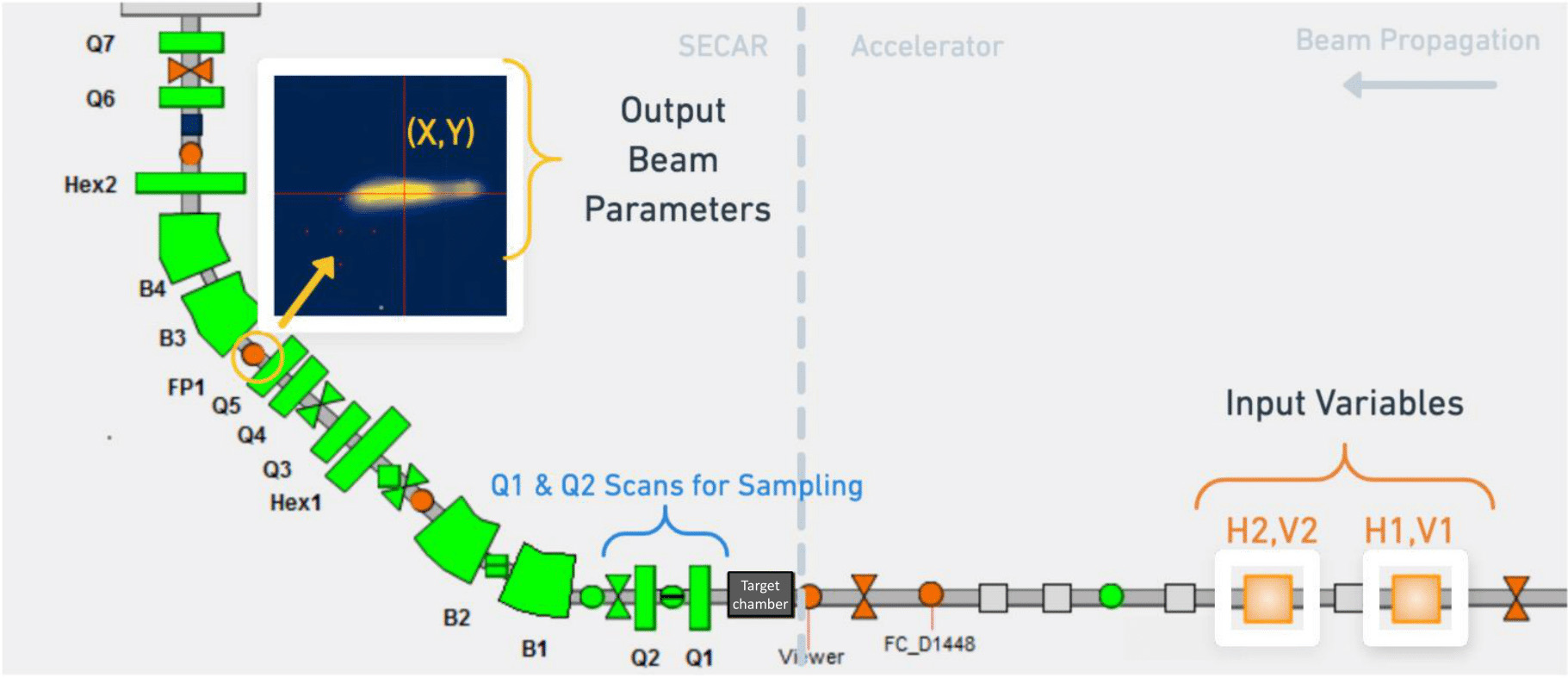}}

\subfloat[][Sampling an observation at the viewer at FP1. The mean steering distance between the four beam spots is 4.7 mm (X and Y in pixels, 1 pixels is $\sim$ 0.31 mm).]{\includegraphics[height=0.17\textheight]{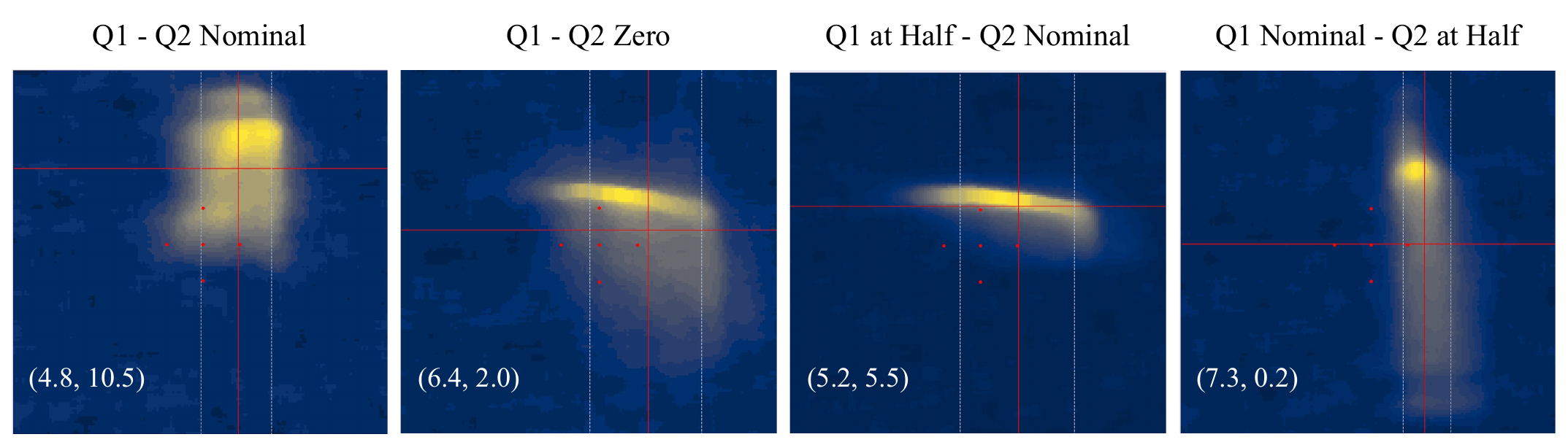}}
\quad
    \caption{Experimental setup for Bayesian optimization of the incoming beam angular deviation.  }
    \label{fig:layout}
    \end{figure}
    
In the absence of a direct way to simultaneously measure and minimize the incoming beam angle at the SECAR target, an indirect method was used employing existing beamline electromagnetic elements and diagnostics. When the incoming beam is deviated from the optical axis of a quadrupole magnet, the beam experiences imbalanced forces and deviates (steers) from its incoming axis when the quadrupole magnet strength is varied. A perfectly tuned beam in SECAR should be transmitted through the optical axis of all the magnets. Based on this behavior, a tuning procedure of the two sets of electromagnetic horizontal (H1, H2) and vertical steerers (V1, V2) installed upstream of SECAR (shown in Figure \ref{fig:layout}) was developed using the SECAR Q1 and Q2 quad\-ru\-poles. By selecting a suitable setting for the two sets of upstream steerers that minimizes the steering produced by Q1 and Q2, a beam with an optimal incoming beam angle and transmission into SECAR was achieved. A single number that quantifies the steering as a function of steerer strengths was obtained using digitized images from the viewer at the first ion optical focus location (FP1 in Figure \ref{fig:layout}). This was done by measuring the center location of the beam spot (X and Y in reference to the viewer center) at four different Q1 and Q2 field strengths, and calculating the mean distance between the four center locations. An example of this sampling is shown in Figure \ref{fig:layout} (b). 
 
\paragraph{Bayesian optimization}
Bayesian optimization is an iterative search for a better optimum that imposes a probabilistic distribution over the objective function values \cite{Kushner1964ANM,Mockus1975,brochu2010tutorial}. It utilizes an acquisition function that places the criterion on how to select the next point based on the probability distribution derived from the observed data, as well as any prior information available. In this work, we model the beam steering in SECAR as a function of steerer magnet settings using a Bayesian approach, and employing quadrupole magnets to obtain the objective function $f$ at the viewer at FP1 (see Figure \ref{fig:layout}). The model takes the steerer magnet current as inputs $(H1, V1, H2, V2)$ and provides an estimate $f$ with a mean $\mu$ and standard deviation $\sigma$ of the steering at FP1 due to quadrupole magnets.

\paragraph{GP}
We chose the popular Gaussian process (GP) \cite{Rasmussen2018} as the probabilistic model as it is flexible, easy to implement, and conveniently provides uncertainty estimates along with its predictions. We describe the beam's response to changes in the settings of different magnetic elements in the beamline by a squared exponential (SE) kernel. The SE function is infinitely differentiable and thus is very smooth, providing a good model for beam response to magnetic field changes. In this work, a prior of a Gaussian form was initialized for the lengthscale $\ell$ and noise $\sigma_n$ SE kernel hyperparameters with a mean and variance based on past tuning experience. The covariance function learns the kernel hyperparameters empirically and updates the priors through maximizing the log marginal likelihood at each step. 

\paragraph{Acquisition function}
Bayesian optimization utilizes the observed data to decide where to evaluate the objective function at each step guided by a lower confidence bound (LCB) acquisition function \cite{Cox97}. It is constructed from the GP posterior mean function $\mu$(x) and its standard deviation $\sigma$(x) as $LCB(x) = \mu(x) - \xi \, \sigma(x)$. The user defined exploration weight $\xi$ directly balances the trade-off between exploiting regions of low mean and exploring regions areas of large uncertainty. 

\paragraph{Algorithm} At initialization, the steering (evaluated using different Q1 and Q2 settings) from a random selection of initial steerer inputs $(H1, V1, H2, V2)$ is evaluated. Next, $x^* = argmin (LCB(x))$ is computed, and these new steerer inputs $x^*$ are used to obtain the next evaluation of the steering $f$. This new observation is added to the samples, and the sequence iterates until the steering is minimized and the beam is fully transmitted. 

\paragraph{Implementation}\label{implementation} This work was implemented using the Python GPy library \cite{GPy} and the associated GPyOpt tool \cite{Authors2016} (BSD license), and integrated into the SECAR control system using PyEpics, a Python interface to the EPICS Channel Access (CA) library for the EPICS control system \cite{Newville_2017}.

\section{Results from SECAR beam commissioning}

The presented work was developed during initial SECAR stable beam commissioning as part of the development of beam tuning procedures and optimizations for the device. Beam species included \textsuperscript{2}H\textsuperscript{1+}, \textsuperscript{133}Cs\textsuperscript{41+}, and \textsuperscript{20}Ne\textsuperscript{8+}, spanning a magnetic rigidity range of 0.1444 to 0.4667 Tm. 

\begin{figure}
\centering
\subfloat[][Horizontal optimization, $\ell$ = 10.8 A, $\sigma_{n}$ = 0.3 mm, $\xi$ = 3]{\includegraphics[scale=0.27]{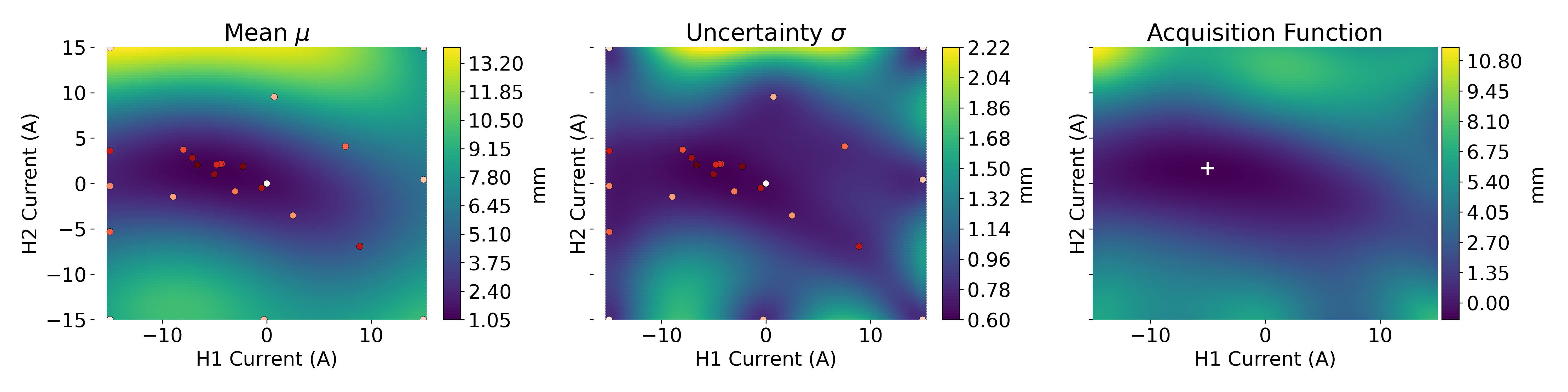}}

\subfloat[][Vertical optimization, $\ell$ = 1.2 A, $\sigma_{n}$ = 0.2 mm, $\xi$ = 0.5]{ \includegraphics[scale=0.27]{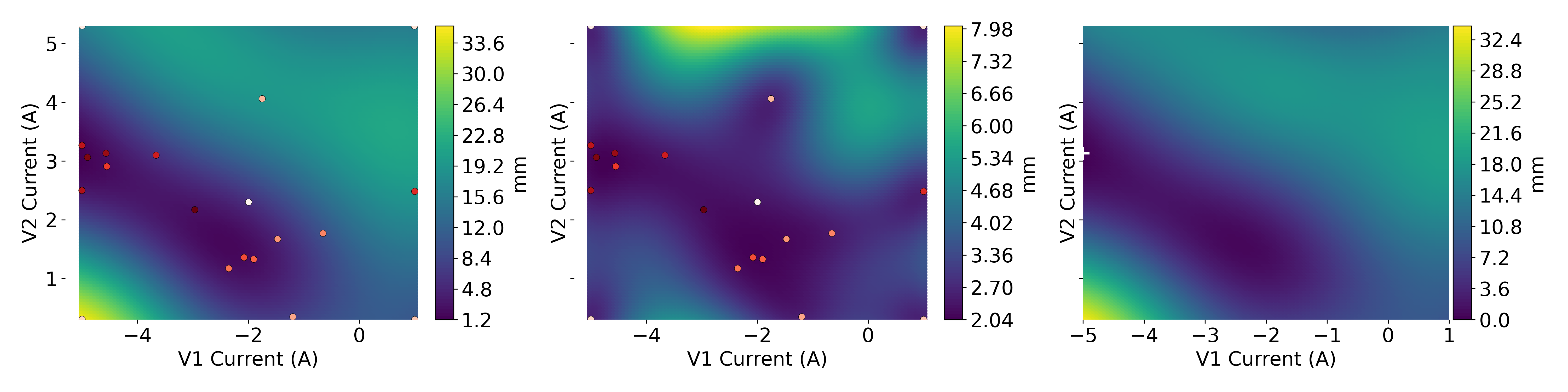}}
\quad
    \caption{The posterior mean $\mu$, standard deviation of the mean $\sigma$, and LCB function for two optimization runs. The observations are shown by dots shaded by time of observation, with the darkest shade being the most recent observation. The next sample point is indicated by the white cross indicating the minimum of the LCB acquisition function. }
    \label{fig:2d results}
    \end{figure}

A typical 2D steerer optimization (two of the steerers were kept at fixed values) is shown in Figure \ref{fig:2d results}. Observations are shown as circular data points (starting with white and increasing in shade up to the most recent observation in red) overlayed on the mean and uncertainty standard deviation plots. The hyperparameters were initialized with priors of $\ell_{prior} = \mathcal{N}(10, 2)$ (A) and $\sigma_{n, prior} = \mathcal{N}(0.31, 0.22)$ (mm) in Figure \ref{fig:2d results} (a) and $\ell_{prior} = \mathcal{N}(2, 1)$ (A) and $\sigma_{n, prior} = \mathcal{N}(0.21, 0.08)$ (mm) for Figure \ref{fig:2d results} (b). 
\begin{figure}
    \centering
    \includegraphics[scale=0.3]{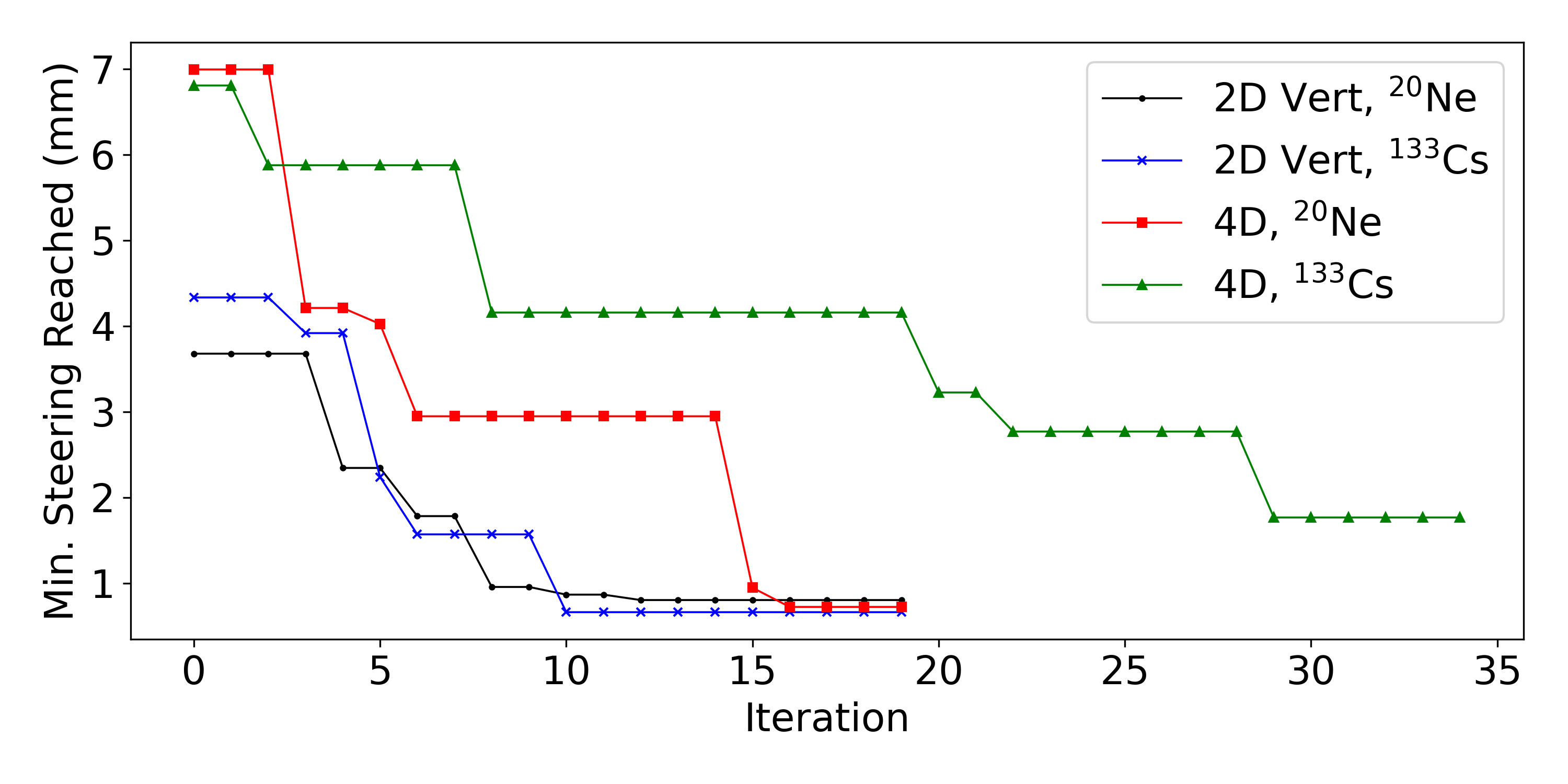}
    \caption{Best steering distance reached as a function of number of GP iterations, for 2D and 4D optimizations during two different beam commissioning runs.}
    \label{fig:steererconv}
\end{figure}

Figure \ref{fig:steererconv} shows the number of iterations to obtain convergence of the algorithm for 2D and 4D optimizations with different beams. The domain of possible steerer values varied between (-3, 3) A up to (-10, 10) A for the 2D and 4D cases, respectively. In general, all 2D optimizations converged within 15 to 20 iterations, while 4D optimizations generally took about double the number of iterations, occasionally needing around 60 when the tune was particularly difficult. At each iteration, adjusting of the quadrupoles takes up to 10 s until they settle. Thus, typical optimization times range between 20 to 40 minutes, which is a significant improvement over the total time spent manually tuning (at least 1-2 hours). The method was robust to changes in the initial conditions of the beam, and sensitive to the accelerator tune upstream. Large variations (large drift, energy re-tunes, hysteresis etc) in settings upstream can affect the final configuration for the steerers, but generally recovers reproducibility along the SECAR as we realign to the beam axis.

While the amount of steering indicates how the angular deviation of the beam from the optical axis is changing, quantifying the incoming angle is not straightforward. To that end, the COSY INFINITY beam physics model \cite{MAKINO2006346} was used to model the incoming angle that could create the observed level of steering for several runs. This was achieved by comparing beam spot locations in COSY to those obtained experimentally with each quad\-ru\-pole tune used to calculate the steering. We found that the incoming angle could typically be reduced to 0.8 mrad after optimization based on the amount of final steering observed experimentally. These results highlight the ability of this method to ensure angles of <1 mrad that are needed for optimal SECAR performance. 

The method shown here for the first section was applied to several similar combinations of quad\-ru\-poles and downstream viewers along the SECAR beamline. This was especially helpful to adjust small angular deviations whose effects were negligible at upstream viewer locations, but significant towards the last section of SECAR. 

\section{Conclusion}
In this paper, we present the first development of online Bayesian optimization for tuning an ion beam in a nuclear astrophysics recoil separator. The method increases the efficiency in achieving the stringent requirements needed for optimal separator performance by at least 3 times as compared to traditional manual tuning methods. We showed that an incoming beam angular deviation was minimized within the specified requirements down to angles of 0-1 mrad. This method is now used routinely for all separator tuning, and can generally be applied to other similar beamlines.

Several improvements on the current optimization are being explored. For instance, the extensive database of accelerator historical data may be used in implementing physics-informed optimizations of incoming beam parameters. Additionally, beam specific priors can be developed by establishing a relationship between beam species, beam rigidity, and the GP kernel hyperparameters. More generally, Bayesian optimization can be used to target other system parameters. One such target is the beam rejection of the separator that can be improved by minimizing the beam size at the mass separation focal planes. This was explored during commissioning as well using online image-based Bayesian optimization of the ion optical system, and is being further studied for use ahead of future scientific experiments.

\section*{Impact statement}
Recoil separators play an important role in the nuclear astrophysics community where there is a need for precise capture reaction rate measurements to answer many open questions relevant to a breadth of stellar nucleosynthesis sites \cite{Ruiz2014}. At astrophysical energies where the yields are low, it becomes crucial to optimize the separators to maximize the rejection of unwanted beam particles. The optimization of such complex high dimensional systems currently requires experienced operators that are forced to balance the time spent on manual tuning, and the time spent on scientific data collection. We presented an online Bayesian optimization method that achieves precise beam angular deviation properties in an efficient and objective manner based on ion optical beam behavior. This method can be extended to other recoil separators or similar beamlines that can benefit from automated online tuning. While beam alignment is a common problem in nuclear and accelerator instruments, this approach is applicable with different targets (e.g. ion optical properties) that can be optimized with similar image-based beam analysis or potentially other diagnostics (Faraday cups, position monitors, etc). In the broader accelerator physics community, this work extends prior developments on electron machines \cite{edelen2018opportunities} to heavy isotope and ion beamlines, which presents additional challenges in control due to complex beam dynamics considerations. These results emphasize the impact such methods can make on increasing scientific output at heavy ion facilities, while supporting reproducible and objective research. Negative ethical or societal impacts are not applicable for this work.

\section*{Acknowledgements and Disclosure of Funding}
This material is based upon work supported by the U.S. Department of Energy, Office of Science, Office of Nuclear Physics under award number DE-SC0014384, and the U.S. National Science Foundation under award numbers PHY-1624942, PHY-1913554, PHY-1102511, and PHY 14-30152 (JINA-CEE).

\bibliography{bib-file}

\end{document}